\documentclass[twocolumn,prl,10pt,aps,floatfix]{revtex4-1}
\usepackage{enumerate}
\usepackage{graphicx}
\usepackage{amsmath}
\usepackage{array}
\usepackage{caption}
\usepackage[colorlinks,linkcolor=blue,citecolor=blue,urlcolor=blue]{hyperref}
\usepackage{color}

\begin{document}
\title{Step conductance and spin selectivity\\ 
in a one dimensional tailored conical magnet}  
\author{X. Zotos}
\affiliation{Department of Physics,
University of Crete, 70013 Heraklion, Greece}
\date{\today}

\begin{abstract}
Using an S-matrix formulation we evaluate the conductance
of a one dimensional free electron gas in double exchange interaction 
with a classical conical  magnet.
We find integer conductance steps depending on the energy window of the 
incoming electrons for conical magnets described by a fictitious magnetic 
field of different orientations and modulated profile. 
The conductance windows, that we attribute to potential or 
diffractive scattering, are characterized by spin selectivity  
depending on the fictitious magnetic field direction and chirality. 
Furthermore, we study the conductance of a conical soliton lattice 
and discuss a rationalization of all the conductance data for an 
incoming electron with arbitrary spin direction in terms of scattering of 
an electron with spin along the conical axis.

\end{abstract}
\maketitle

\section{Introduction}
The phenomenon of Chiral Induced Spin Selectivity (CISS) 
recently attracts considerable interest in diverse scientific fields as
chemistry \cite{ciss}, biology \cite{goehler,guo}, physics and 
magnetic materials research \cite{cheong}. 
In particular, in the field of spintronics CISS is promoted as a 
promising mechanism for the generation and control  of spin currents 
\cite{advances_2024,advances_2025}. 
The theory of chiral metallic structures has a long history \cite{jetp}. 
On the bulk level, metallic magnetic systems are often described 
by the so called double exchange interaction model in which 
a lattice of classical spins interacts with the 
conduction electrons through the Hund's rule coupling that
aligns the spins of the conduction band and localized electrons 
occupying the same lattice site. 
The double exchange interaction was shown to lead to complex magnetic phases 
\cite{mostovoy}. In chiral magnetic compounds, 
the spin-transfer torque on a conical magnet 
has been studied \cite{stiles}, nonreciprocal electrical transport observed 
\cite{aoki,inui} and control of the chirality in helimagnets 
by a charge current has been sought \cite{jiang,ohe,masuda}.
Concerning the interaction of a spin-1/2 particle with an helical 
magnetic structure, it has earlier on been studied in the framework 
of neutron scattering in conical magnets \cite{calvo}.   

In this work we study the conductance of a one dimensional electron gas
interacting via a double exchange interaction with a classical conical magnet.
We use an S-matrix formulation \cite{landauer,buttiker} 
that implies coherent electronic transport. 
In particular, (i) we study conical magnets  
with different orientation between the axis of the magnet and 
direction of electron propagation, 
(ii) we introduce magnets with variable radius in order to 
minimize the usual oscillations due to a sharp edge potential.
We put into evidence energy windows in the conductance with sharp steps 
as a function of energy. 
The modulated geometry of the magnet allows us to delineate two distinct 
mechanism of scattering, a potential and a diffractive one.
Each conductance profile has characteristic magnetic field 
direction and chirality dependence. 
We discuss them in terms of the energy spectra of the conical magnets and 
symmetry arguments.

\section{Model and method}
We consider a classical magnetic system interacting 
with a one dimensional 
open bath of free electrons described by the Scr\"odinger equation,

\begin{equation}
H\Psi=
[-\frac{\hbar^2}{2m}\frac{\partial^2}{\partial y^2}+V(y)]\Psi=\epsilon \Psi.
\end{equation}

\noindent
$\Psi$ 
is a two component plane wave wavefunction of wavevector $q$ 
for the $\hat z$-projection of the electron spin 
and $V(y)$ a double exchange interaction,

\begin{equation}
V(y)=-\vec {h}(y) \cdot \vec \sigma,
\label{vx}
\end{equation}

\noindent
${\vec \sigma}$ the spin-1/2 Pauli matrices.
The fictitious magnetic field ${\vec h}$ 
is of the form ${\vec h}(y)=h(y){\hat h(y)}~ (0\le y \le L)$, 
${\hat h}(y)$ is a unit field vector and ${\vec h}(y)=0$ for $y < 0,~ y > L$. 
This field term represents the double exchange interaction,
$-J\frac{1}{2}{\vec \sigma}(y)\cdot {\vec S}(y)$
where $J$ is the strength of the Hund's rule coupling between the 
classical spin $\vec S$ of unit length and the conduction electron 
spin at the same position \cite{mostovoy}. 

We evaluate the Landauer-B\"uttiker conductance $G$ of the electron gas,
employing a  multichannel $S$-matrix formalism\cite{landauer,buttiker},
\begin{equation*}
{\bf S}=
\begin{pmatrix}
r_{\uparrow\uparrow}
&r_{\uparrow\downarrow}
&{\bar t}_{\uparrow\uparrow}
&{\bar t}_{\uparrow\downarrow}\\
r_{\downarrow\uparrow}
&r_{\downarrow\downarrow}
&{\bar t}_{\downarrow\uparrow}
&{\bar t}_{\downarrow\downarrow}\\
t_{\uparrow\uparrow}
&t_{\uparrow\downarrow}
&{\bar r}_{\uparrow\uparrow}
&{\bar r}_{\uparrow\downarrow}\\
t_{\downarrow\uparrow}
&t_{\downarrow\downarrow}
&{\bar r}_{\downarrow\uparrow}
&{\bar r}_{\downarrow\downarrow}
\end{pmatrix}.
\end{equation*}
\noindent
The transmission through the region where the potential (\ref{vx}) is nonzero, 
that is $0<y<L$, is described by the scattering matrix, 
so one considers here the incoming/outgoing states in the leads.
$t_{\sigma',\sigma}$ is the transmission coefficient 
of a spin-$\sigma$ electron 
in the left channel to  a spin-$\sigma'$ in the right channel and 
$r_{\sigma',\sigma}$ the reflection coefficient in the left channel. 
${\bar t}_{\sigma',\sigma},~{\bar r}_{\sigma',\sigma}$ are the corresponing 
coefficients in the right channel. In the following, we consider 
electron propagation along the ${\hat y}$ direction.

The conductance is given by,
\begin{equation}
G=\frac{e^2}{h}\int_0^{\infty} d\epsilon'
(-\frac{\partial f}{\partial \epsilon'} )
\sum_{\sigma'\sigma} |t_{\sigma',\sigma}|^2
\label{g}
\end{equation}
\noindent
with $f(\epsilon)$ the Fermi function,
\begin{equation*}
f(\epsilon)=\frac{1}{1+e^{\beta(\epsilon-\mu)}},
\end{equation*}
$\beta=1/k_BT$, $T$ the temperature and $\mu$ the chemical potential.
We take $\hbar^2/2m=1$ so that $\epsilon=q^2$ and $k_B=1$. 
We consider the zero temperature limit,
$T\rightarrow 0~(\beta\rightarrow \infty)$ where the term 
$-\partial f/\partial \epsilon'$ tends to a $\delta$-function at the 
chemical potential $\mu$ 
(from here on we take the unit of conductance, $e^2/\hbar=1$).
We thus present the conductance $G$ 
as a function of energy $\epsilon(=\mu)$, in order to make 
clear the correspondence of $G$ to the energy spectrum, Fig.\ref{fig3}, 
of the conical magnets as discussed below.
As expected, a finite temperature would smooth the conductance curves.

We construct the $\bf S$ matrix of various geometries 
conical magnets by decomposing the interval $L$ in slices 
of width $dx$ and by successive ${\hat x},{\hat y},{\hat z}$-dependent
rotations,
\begin{eqnarray*}
R_z(\phi)&=&
\begin{pmatrix}
e^{i\phi/2}&0\\
0&e^{-i\phi/2}\\
\end{pmatrix},
\nonumber\\
R_x(\phi)&=&
\begin{pmatrix}
\cos(\phi/2)&i\sin(\phi/2)\\
i\sin(\phi/2)&\cos(\phi/2)
\end{pmatrix},
\nonumber\\
R_y(\theta)&=&
\begin{pmatrix}
+\cos(\phi/2)&+\sin(\phi/2)\\
-\sin(\phi/2)&+\cos(\phi/2)
\end{pmatrix},
\end{eqnarray*}
\noindent
of the Hamiltonian that make the scattering diagonal in each slice
(we have also verified the results by a T-matrix approach, which however 
is unstable for large systems due to the apperance of exponentially 
growing factors). 

\section{ ${\hat y}$ - conical magnet}
\begin{figure}[ht]
\includegraphics[angle=0, width=0.8\linewidth]{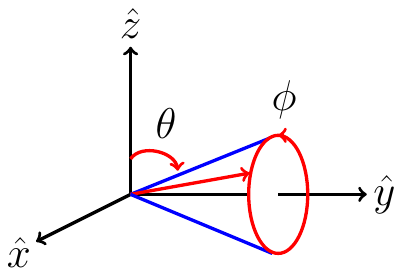}
\caption{${\hat y}$ - conical magnet.}
\label{fig1}
\end{figure}

We first create a conical magnet in the ${\hat y}$ - direction, 
Fig.\ref{fig1}, 
by a rotation $R_x(\theta)$ by an angle $\theta$
of the unit field vector from the $+{\hat z}$ direction, followed 
by a rotation $R_y(\phi)$, where $\phi(y)=Qy$. 
$Q$ is the characteristic wavevector so that the magnetic field is 
given by, ${\vec h}=h{\hat h}= 
h (\cos \theta\sin Qy,\sin \theta,\cos \theta \cos Qy)$.

\begin{figure}[ht]
\includegraphics[angle=0, width=0.9\linewidth]{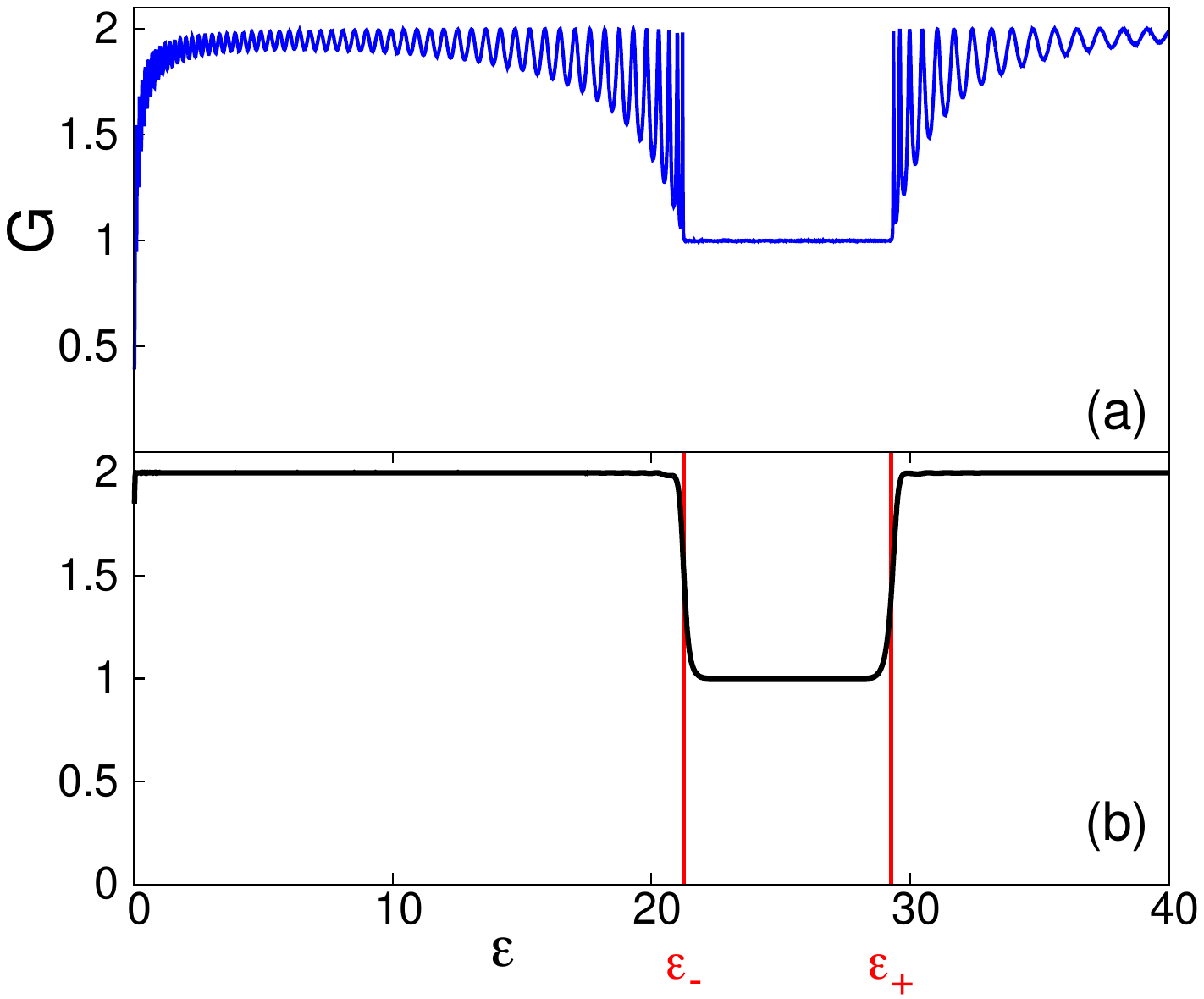}
\caption{Conductance as a function of energy of a spiral magnet rotating 
around the ${\hat y}$-axis,
(a) ${\vec h}(y)=h(\cos Qy,0,\sin Qy)$, 
(b) ${\vec h}(y)=h\sin(\pi x/L)(\cos Qy,0,\sin Qy)$, 
$\epsilon_{\pm}=(Q/2)^2\pm h$ (red).}
\label{fig2}
\end{figure}

In Fig.\ref{fig2}a we show the total conductance $G$ as a function of 
the energy $\epsilon$ of incoming electrons.
Here, we consider a ''spiral" magnet where $\theta=0$, the unit 
field vector rotating in the ${\hat x}-{\hat z}$ plane, 
${\vec h}(y) =h(\sin Qy,0,\cos Qy)$.
Hereafter, we take $L=40~, h=4$ and $Q=2\pi m/L~ (m=64)$.
The main result is that the conductance takes the value $G=1$ in a 
window of energies $\epsilon_- < \epsilon < \epsilon_+,~ 
\epsilon_{\pm}=(Q/2)^2\pm h$ that we will call {\it diffraction} 
window as it is related to the wavevector $Q$ of the spiral magnet.

To supress the usual conductance oscillations due to the sharp edges of the 
magnetic field potential, we superimpose a smooth profile so that  
${\vec h}(y)=h\sin(\pi y/L){\hat h}(y)$ \cite{ssd}.
In Fig.\ref{fig2}b we find a step like conductance with $G=1$ in the same 
window as above (actually tending to a step by increasing the 
length of the system, $L\rightarrow \infty$) and $G=2$ outside the 
diffraction window. Hereafter, the quoted integer values of the 
conductance we find are accurate to order 10$^{-4}$.    

\begin{figure}[ht]
\includegraphics[angle=0, width=0.9\linewidth]{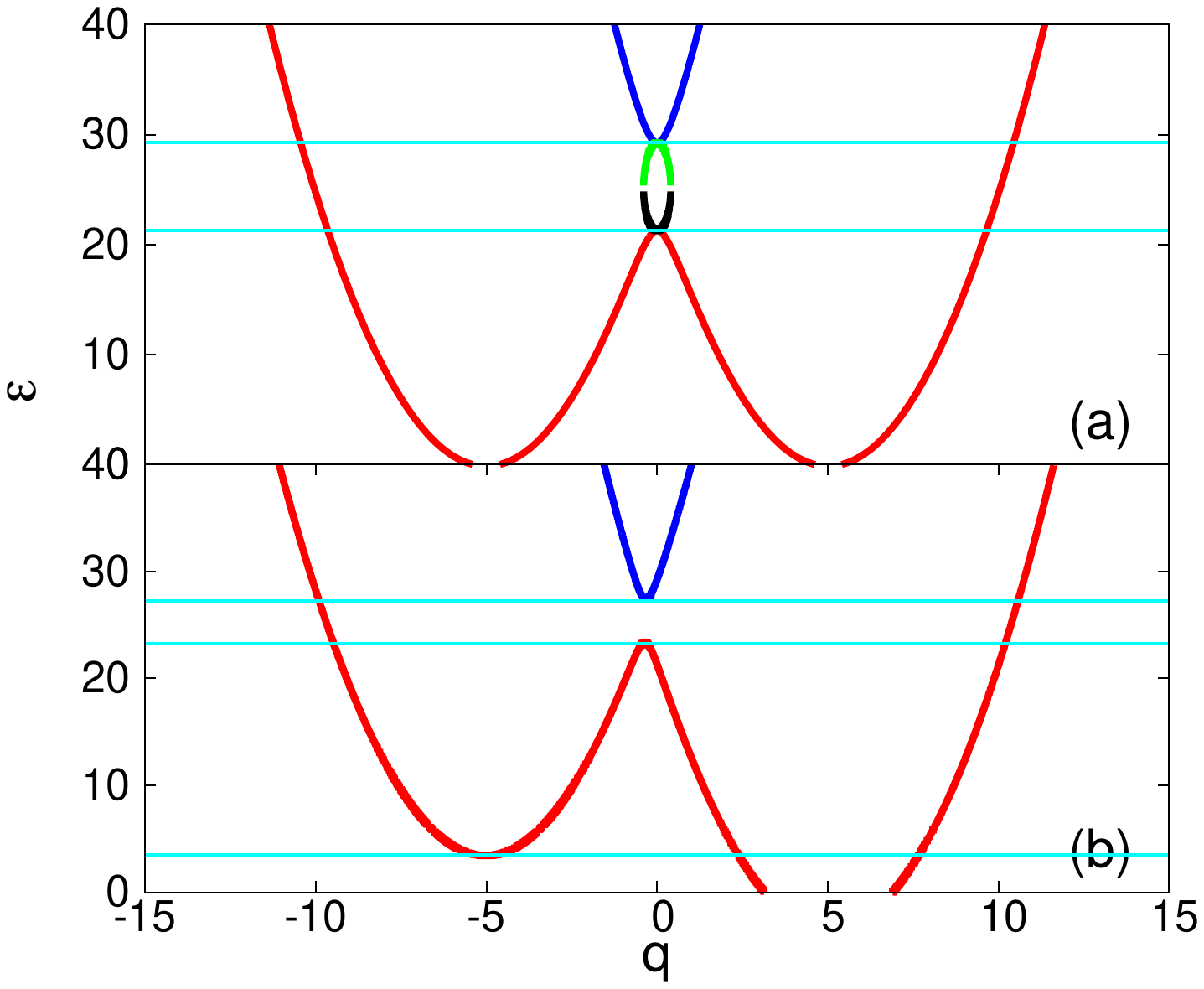}
\caption{Spectrum of the ${\hat y}$ - conical  magnet, 
(a) $\theta=0$, $\epsilon_{\pm}=
(Q/2)^2\pm h$ (cyan),
$\epsilon_-(q)$ (blue), $\epsilon_+(q)$ (red), 
${\bar \epsilon}_-(iq)$ (black),
${\bar \epsilon}_+(iq)$ (green), 
(b) $\theta=\pi/3$, $\epsilon_0=h\sin \theta$, $\epsilon_{\pm}=
(Q/2)^2\pm h\cos \theta$ (cyan).}
\label{fig3}
\end{figure}

To get insight on the $G=1$ window in the conductance, we diagonalize the 
Hamiltonian of the conical magnet \cite{calvo} 
by a rotation of the wavefunction, 
\begin{eqnarray*}
&&|\Psi_{\pm}>=e^{iqy}R^{-1}_y(\phi(y))\cdot 
R^{-1}_x(w(\theta,q))\cdot \eta_{\pm},
\nonumber\\
&&\eta_+=
\begin{bmatrix}
1\\
0
\end{bmatrix},~~
\eta_-=
\begin{bmatrix}
0\\
1
\end{bmatrix},
\nonumber\\
&&R_y(\phi(y))\cdot H \cdot R^{-1}_y(\phi(y))=
\nonumber\\
&&\Big(-\frac{\partial^2}{\partial y^2} +(Q/2)^2
+(i Q \frac{\partial}{\partial y} - h \sin \theta)
 \sigma^y -h \cos \theta \sigma^z\Big),
\end{eqnarray*}
\begin{eqnarray*}
\tan w/2&=&\frac{h\sin \theta+qQ}{\sqrt{h^2+(qQ)^2+2qQh\sin \theta}
-h\cos \theta},
\end{eqnarray*}
\noindent
and obtain the spectrum,
\begin{equation*}
\epsilon_{\pm}=q^2+(Q/2)^2\pm \sqrt{ h^2+(qQ)^2 +2qQh\sin \theta}.
\end{equation*}
\noindent
As shown in Fig.\ref{fig3}a,
for the spiral magnet ($\theta=0$) there are four energy branches:
\begin{eqnarray*}
\epsilon_{\pm}&=&q^2+(Q/2)^2\pm \sqrt{h^2+(qQ)^2}
\nonumber\\
{\bar \epsilon}_{\pm}
&=&-{\bar q}^2+(Q/2)^2\pm \sqrt{h^2-({\bar q}Q)^2},~~q=i{\bar q}
\end{eqnarray*}
\noindent
and in particular two with imaginary wavevector $q$ 
for energy $\epsilon_- < \epsilon < \epsilon_+$ that do not contribute 
to the conductance, thus $G=1$. 
Note that the scattering 
coefficients can be evaluated semi-analytically 
by the inversion of an 8x8 matrix, obtained from  the continuity 
of the wavefunction and its 1st derivative. 

For the partial transmissions composing the total conductance $G$ 
in Fig.\ref{fig2}b we find by,

\noindent
invariance under rotation around the ${\hat y}$ axis,  
\begin{equation}
|t_{\sigma,\sigma}|^2=|t_{-\sigma,-\sigma}|^2,~~
|t_{\sigma,-\sigma}|^2=|t_{-\sigma,\sigma}|^2,
\label{updown}
\end{equation}
\noindent
inversion of the magnetic field direction, 
\begin{equation}
|t_{\sigma',\sigma}(+h)|^2=|t_{\sigma',\sigma}(-h)|^2,
\label{hmh}
\end{equation}
\noindent
inversion of propagation direction (reciprocity),
\begin{equation}
|t_{\sigma',\sigma}|^2=|{\bar t}_{\sigma',\sigma}|^2,
\label{lr}
\end{equation}
\noindent
and inversion of magnetic field chirality,
\begin{equation}
|t_{\sigma',\sigma}(+Q)|^2=|t_{\sigma',\sigma}(-Q)|^2.
\label{inversion}
\end{equation}
\noindent
We should emphasize that the equalities in these relations 
are in the large $L$ limit and the effect of the 
superimposed smooth profile that suppresses reflection oscillations. 

\begin{figure}[ht]
\includegraphics[angle=0, width=0.9\linewidth]{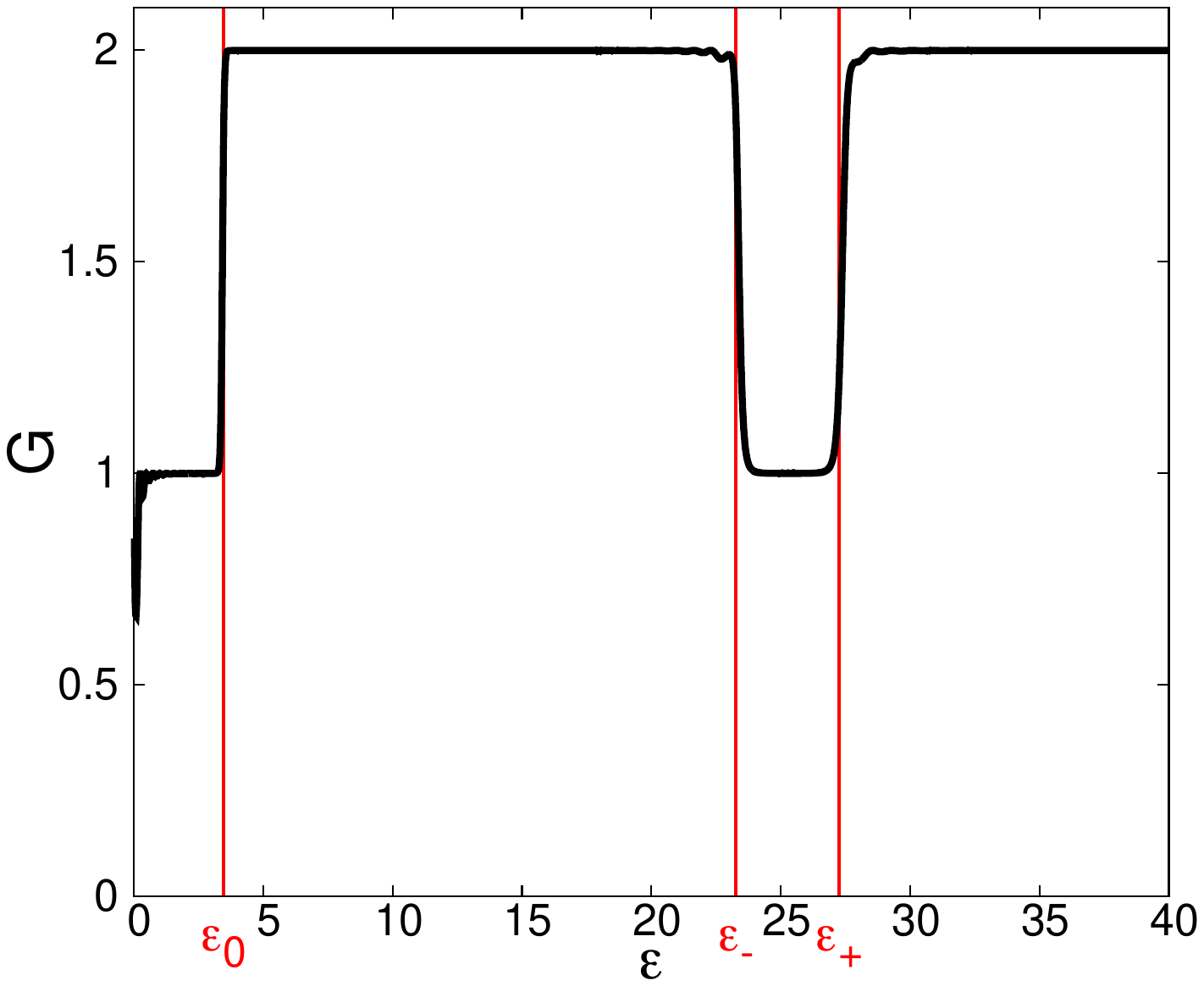}
\caption{Conductance as a function of energy 
of a modulated ${\hat y}$-conical magnet $(\theta=\pi/3)$, 
$\epsilon_{\pm}=
(Q/2)^2\pm h\cos\theta,~ \epsilon_0=h\sin \theta$.}
\label{fig4}
\end{figure}

Next, in Fig.\ref{fig4} we show the conductance of a modulated 
${\hat y}$ - conical magnet for $\theta=\pi/3$, ${\vec h}(y)=h\sin(\pi y/L)
(\cos \theta\sin Qy,\sin \theta,\cos \theta\cos Qy)$.
There are two windows of $G=1$ conductance that we attribute to 
different scattering mechanism. 
For $\epsilon < \epsilon_0~ (\epsilon_0= h\sin \theta)$ 
there is {\it potential} scattering 
of the  incoming electrons for energies below the ${\hat y}$ - component 
of the magnetic field. 
For $\epsilon_- < \epsilon < \epsilon_+
~ (\epsilon_{\pm}=(Q/2)^2\pm h \cos \theta)$ 
there is a diffraction window, the same as in the spiral magnet, but with 
boundary magnetic fields the projection in the ${\hat z}$ - direction.
The suppression of the conductance to $G=1$ in these windows is consistent 
with the energy dispersions in the  spectrum shown in Fig.\ref{fig3}b. 

In the potential window, $\epsilon < \epsilon_0$, 
relations (\ref{updown},\ref{hmh},\ref{lr},\ref{inversion}) 
for the partial transmissions still hold. 
These values are consistent with potential scattering of a 
spin $\sigma=\uparrow,\downarrow$ electron 
projected along the ${\hat y}$ - direction.

In the diffraction window, $\epsilon_- < \epsilon < \epsilon_+$,  
due to the invariance under ${\hat y}$ - axis rotation, still holds,
\begin{equation*}
|t_{\uparrow\uparrow}|^2=|t_{\downarrow\downarrow}|^2,~~
|t_{\uparrow\downarrow}|^2=|t_{\downarrow\uparrow}|^2.
\end{equation*}
\noindent
However, due to the ${\hat y}$ directional magnetic field 
breaking, by reversing the magnetic field $h$, wavevector $Q$ or 
propagation direction, the partial  transmissions
(\ref{inversion}) are replaced by,
\begin{equation*}
|t_{\sigma\sigma'}(+h,+Q)|^2=
|{\bar t}_{\sigma\sigma'}(-h,+Q)|^2=
|{\bar t}_{\sigma'\sigma'}(+h,-Q)|^2.
\end{equation*}
\noindent
In the $G=1$ potential and diffraction conductance windows, the 
transmission and reflection coefficients are equal to 
$|t_{\sigma,\sigma}|^2=|t_{\sigma,-\sigma}|^2=1/4,~~
|r_{\sigma,\sigma}|^2=|r_{\sigma,-\sigma}|^2=1/4$. For energies outside these 
windows the reflection coefficients vanish.
The above results will be rationalized in the following section on 
{\bf Spin selectivity}.

\begin{figure}[ht]
\includegraphics[angle=0, width=0.8\linewidth] {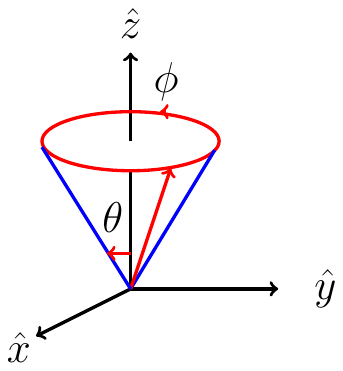}
\caption{$\hat z$ - conical magnet.}
\label{fig5}
\end{figure}

\begin{figure}[ht]
\includegraphics[angle=0, width=0.9\linewidth]{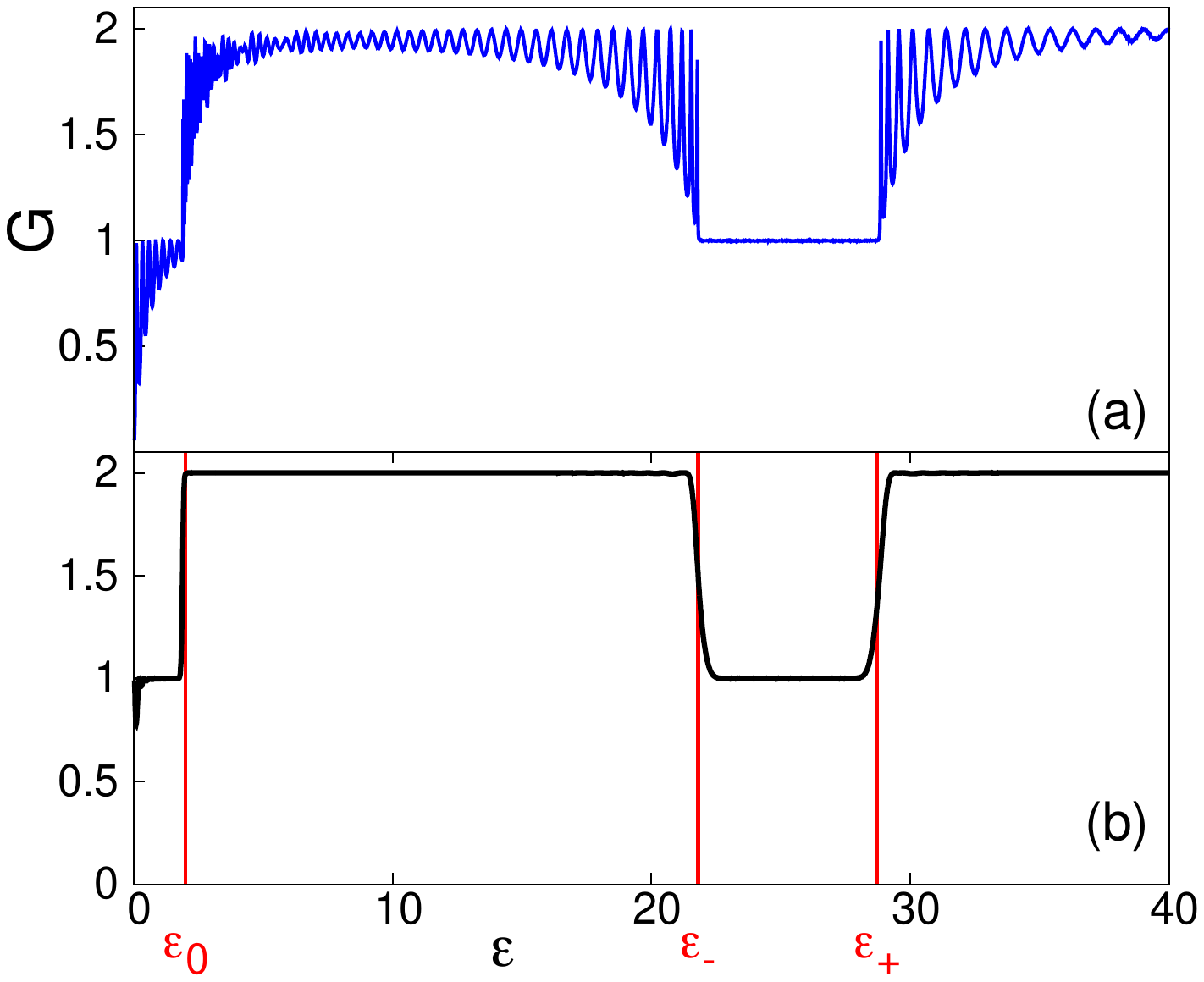}
\caption{Conductance $G$ as a function of energy 
of a ${\hat z}$-conical magnet $(\theta=\pi/3)$, 
$\epsilon_{\pm}=(Q/2)^2\pm h\sin \theta,~ \epsilon_0=h\cos \theta$. 
a) 
${\vec h}(x)=h{\hat h}(x)=
h(\sin \theta \cos \phi, \sin \theta \sin \phi,\cos \theta)$, 
b) ${\vec h}(x)=h\sin(\pi x/L){\hat h}(x)$.}
\label{fig6}
\end{figure}

\begin{figure}[ht]
\includegraphics[angle=0, width=0.9\linewidth]{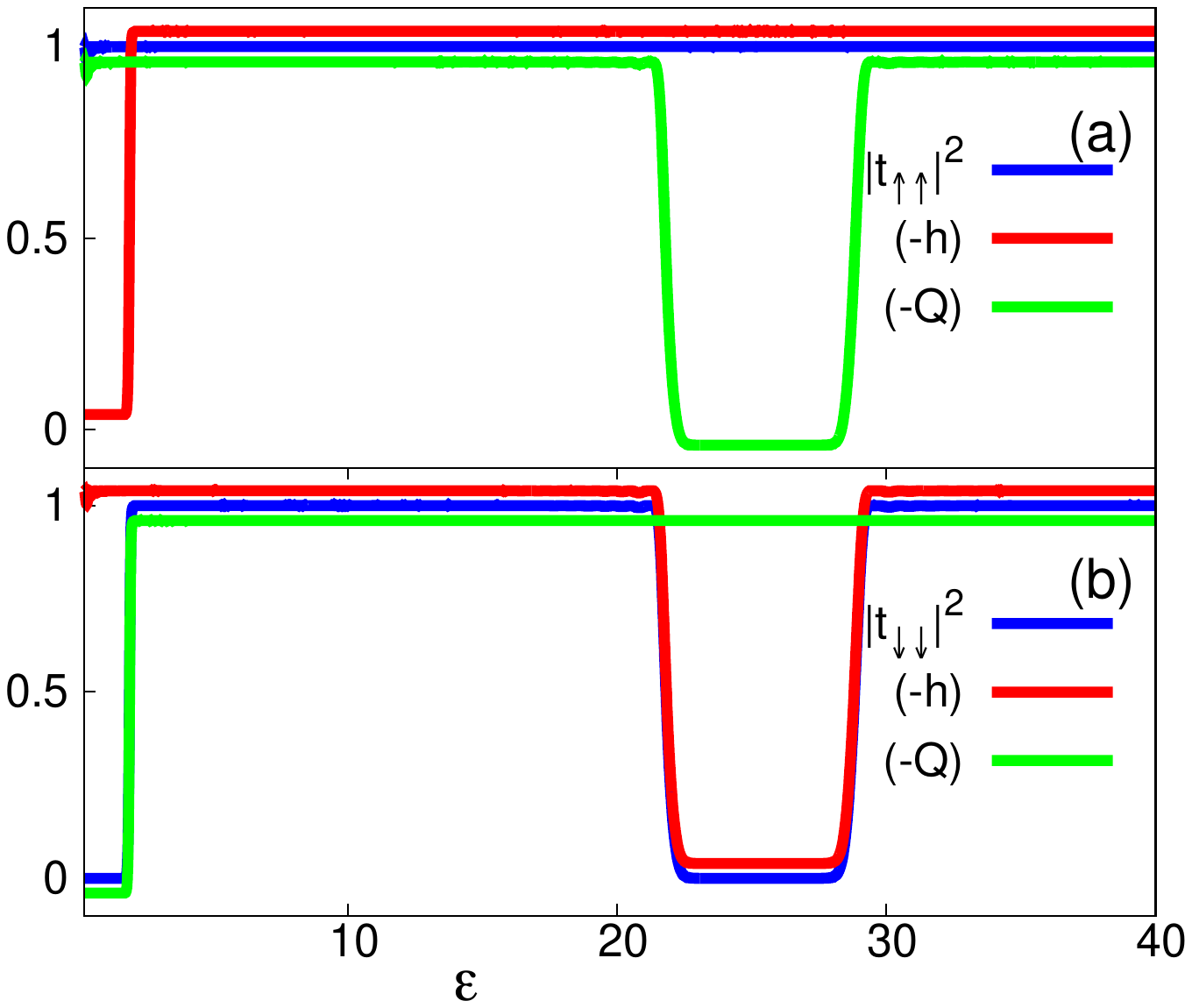}
\caption{Partial conductances (of Fig.\ref{fig6}b) in the forward direction 
as a function of energy 
of a modulated ${\hat z}$-conical magnet 
((-h),(-Q) curves sligthly displaced for clarity).}
\label{fig7}
\end{figure}

\begin{figure}[ht]
\includegraphics[angle=0, width=0.9\linewidth]{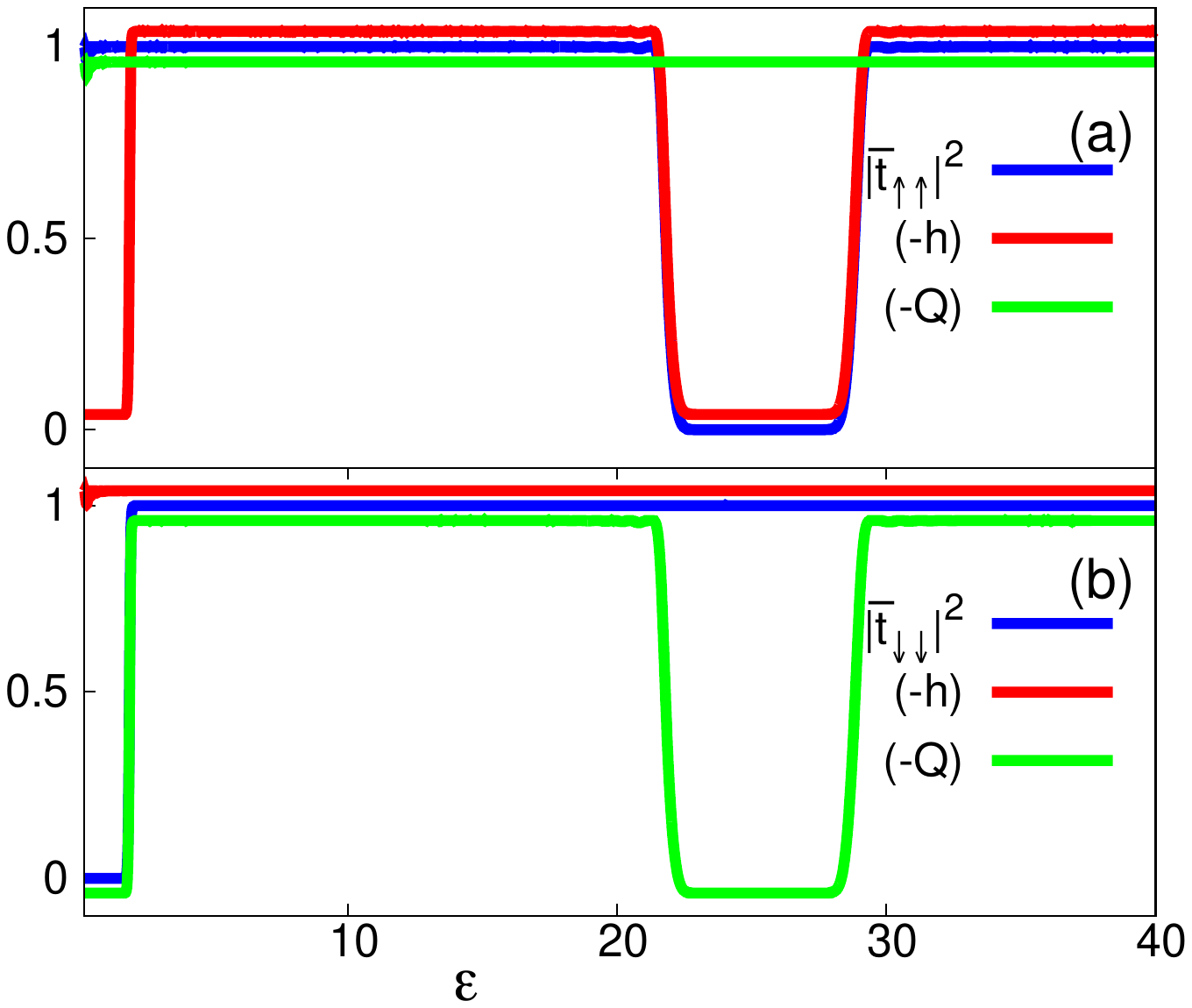}
\caption{Partial conductances (of Fig.\ref{fig6}b) in the backward direction 
as a function of energy $\epsilon$ 
of a modulated ${\hat z}$-conical magnet
((-h),(-Q) curves slightly displaced for clarity).}
\label{fig8}
\end{figure}

\section{$\hat z$ - conical magnet}
Next, we study a conical magnet along the ${\hat z}$-axis, Fig.\ref{fig5}, 
${\vec h}(y)=h(\sin \theta \cos \phi, \sin \theta \sin \phi,\cos \theta)$,
created by a rotation by an angle  $\theta$  around the ${\hat y}$ or 
${\hat x}$ axis, followed by a $\phi=Qy$ rotation around the ${\hat z}$ - axis. 
We can again  diagonalize the Hamiltonian by an axis rotation to obtain 
the eigenstates, 
\begin{equation*}
|\Psi_{\pm}>=e^{iqy}R^{-1}_z(\phi(y))\cdot 
R^{-1}_x(w(\theta,q))\cdot \eta_{\pm},
\end{equation*}
\noindent
and eigenvalues,
\begin{equation*}
\epsilon_{\pm}=q^2+(Q/2)^2\pm \sqrt{ h^2+(qQ)^2 +2qQh\cos \theta}.
\end{equation*}
\noindent
The spectrum is the same as in the ${\hat y}$ - conical magnet, 
with the same dispersion relations and propagating modes.

Again, to eliminate oscillatory behavior of the conductance, we  
modulate the field by a $\sin \pi y/L$ prefactor and obtain  
$G(\epsilon)$ as shown in Fig.\ref{fig6} ((a) unmodulated, (b) modulated).
Here, $G={\bar G}$ and there are two energy windows that 
we will assign, based on Figs.\ref{fig7},\ref{fig8},  
as due to, (i) {\it potential} scattering
for $\epsilon < \epsilon_0$, where $\epsilon_0$ is  equal 
to the ${\hat z}$-component 
of the magnetic field and (ii) {\it diffractive} scattering for
$\epsilon_- < \epsilon < \epsilon_+$ e.g. at energies related 
to the $Q$ wavevector. 
Before proceeding, we should mention again that the step and integer like 
conductance data discussed below, are in the large $L$ limit 
and in the presence of the smoothing potential.

First, we find (not shown) that the off-diagonal,  
spin-flip partial transmissions vanish, 
$|t_{\sigma'\ne \sigma}|^2=|{\bar t}_{\sigma'\ne \sigma}|^2=0$. 
Next, in the first window, $0 < \epsilon < \epsilon_0$, 
the transmission depends 
only on the direction of the spin to the magnetic field, as expected 
from potential scattering and it is independent on the chirality of the 
magnetic field. This holds for the forward as well as backward transmission. 
Also, in Figs.\ref{fig7},\ref{fig8}, in the potential window where the 
partial conductance vanishes, the reflection coefficient is unity, e.g. when
$|t_{\sigma\sigma}|^2=0,~~|r_{\sigma\sigma}|^2=1$, in other words 
the spin direction is conserved in the {\it potential} reflection. 

The second, transmission window, $\epsilon_- < \epsilon < \epsilon_+$, 
depends  on the chirality of the magnetic field.
For instance, as shown in Fig.\ref{fig7}a, 
$|t_{\uparrow\uparrow}|^2=1$ for all energies as an up-spin 
sees a negative potential and positive chirality (blue curve). Inverting 
the magnetic field ($+h\rightarrow -h$) the first window appears (red curve), 
while reversing $Q$ the second window appears (green curve). 
The opposite behavior is shown in Fig.\ref{fig7}b for the down-spin, where
$|t_{\downarrow\downarrow}|^2$ shows two windows, the first suppressed by 
inverting the magnetic field and the second by reversing the magnetic field
chirality. 
Thus $|t_{\downarrow\downarrow}(-h,-Q)|^2=|t_{\uparrow\uparrow}|^2=1$.
This means that there is spin selectivity by reversing the chirality of 
the magnetic field.

At the same time, in the diffraction window where the 
partial conductance vanishes, the spin-flip reflection coefficient equals one, 
e.g. when $|t_{\sigma,\sigma}|^2=0,~~|r_{\sigma,-\sigma}|^2=1$.
In the backward transmissions $|{\bar t}_{\sigma'\sigma}|^2$ 
shown in Fig.\ref{fig8}, 
as expected, the role of the magnetic field remains the same, while that of the 
chirality is reversed.
The above data indicate spin selectivity, depending on the energy window,
direction of magnetic field and its chirality. 

\section{Chiral Soliton Lattice}
\begin{figure}[ht]
\includegraphics[angle=0, width=0.9\linewidth]{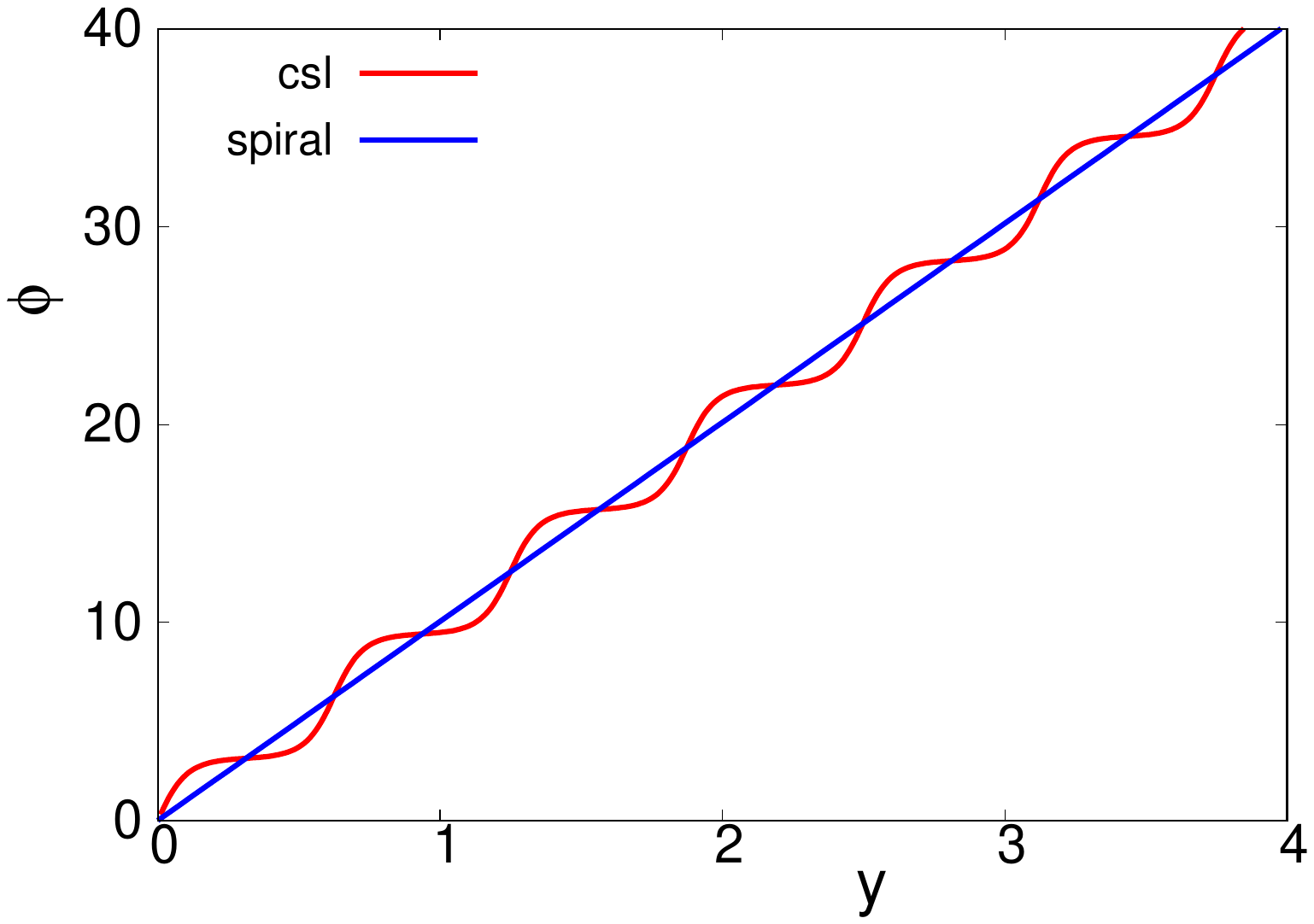}
\caption{Rotation angle in the ${\hat x}$ - ${\hat z}$ - plane 
as a function of distance.} 
\label{fig9}
\end{figure}

The CSL has been studied early on by De Gennes \cite{degennes} 
and is presently in the focus of several experimental studies in a 
variety of compounds, e.g. \cite{csl1,csl2}. 
It appears in a conical magnet under the application of a transverse field,
creating ferromagnetic polarized regions alternating with $2\pi$ solitons.
In the ${\hat y}$ - conical magnet the rotation angle 
in the ${\hat x}$ - ${\hat z}$ plane is deformed, as shown in Fig.\ref{fig9},
under the application of a transverse magnetic field along ${\hat z}$.
The field configuration is given in terms of 
Jacobi elliptic functions, with $m'=0.001$ and $z$ rescaled to $y$ 
so that $Q$ is the same as in the spiral configuration,
${\vec h}=h\sin(\pi y/L)(1-2{\rm sn^2z},0,2{\rm snz~cnz})$.

\begin{figure}[ht]
\includegraphics[angle=0, width=0.9\linewidth]{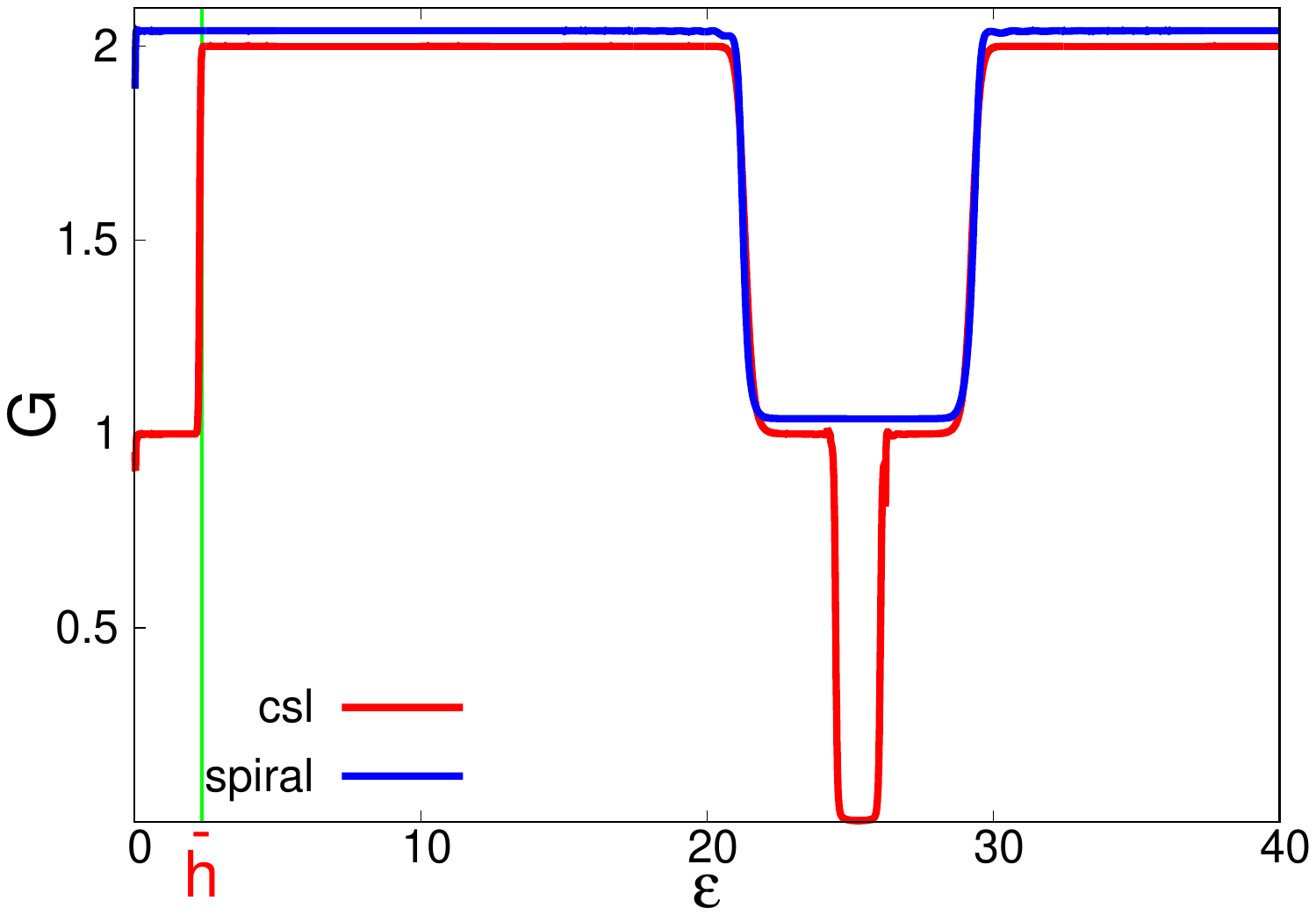}
\caption{Conductance of the chiral soliton lattice as a function of energy. 
${\bar h}$ denotes the average magnetic field due to ferromagnetic regions 
between solitons (blue curve vertically displaced for clarity).}
\label{fig10}
\end{figure}

In Fig.\ref{fig10} we show the conductance for the spiral and the CSL lattice.
In the CSL lattice a potential window appears due to the ferromagnetic 
polarized regions, of width  
${\bar h}=(h/L)\int_0^L dy \cos\phi,~~
\phi=\tan^{-1} 2{\rm snz~cnz} /(1-2{\rm sn^2 z})$.  
In the diffractive window, a region appears of complete reflection, $G=0$, 
with width ${\bar h}$.

\section{Spin selectivity}
To discuss the spin selectivity, we introduce the chirality of the 
magnetic field configuration as,
\begin{equation}
{\vec \chi}={\hat h}\times \frac{\partial {\hat h}}{\partial y}.
\end{equation}
For the ${\hat y}$ - conical magnet shown in Fig.\ref{fig1}, $\chi^y=+Q$, 
while for the ${\hat z}$ - conical shown in Fig.\ref{fig5}, $\chi^z=+Q$.

We can rationalize all the above scattering data in terms of an incoming 
electron with spin parallel or antiparallel to the direction 
of the conical axis.
Let us denote by $|\Uparrow>,~~|\Downarrow>$ 
e.g. the $\pm {\hat y}$ for the ${\hat y}$ - conical 
or the $\pm {\hat z}$ for the ${\hat z}$ - conical magnet basis states.

In the potential window, 
a particle scatters from a positive or negative potential 
box (\ref{vx}), of amplitude the component
of the fictitious magnetic field along the incoming electron spin direction.
The spin flip coefficients vanish, 
$t_{\Uparrow\Downarrow}=t_{\Downarrow\Uparrow}
=r_{\Uparrow\Downarrow}=r_{\Downarrow\Uparrow}=0$.

In the diffraction window, the spin selectivity is given by the 
conditions, (i) an incoming electron with spin parallel to the 
conical axis and the magnetic field 
chirality is fully transmitted, $|t_{\Uparrow\Uparrow}(+Q)|=1$, 
(ii) an electron with spin opposite 
to the chirality is fully reflected with reversal of 
the spin direction, $|r_{\Uparrow\Downarrow}(+Q)|=1$. 

Reversing the chirality, $Q\rightarrow -Q$, clearly 
does not affect the scattering coefficients in the potential window,
while in the diffraction window $|t_{\Downarrow\Downarrow}(-Q)|=1,~ 
|r_{\Downarrow\Uparrow}(-Q)|=1$. 
Reversing the direction of the magnetic field, $h\rightarrow -h$, 
the direction of the transmitted/reflected spin is reversed only in the 
potential window. 
Reversing the direction of propagation from 
right to left, reverses the spin components.  
The ${\hat z}$ - conical magnet results shown in 
Figs.\ref{fig7},\ref{fig8} demonstrate these conditions. 

Considering a left lead incoming electron in an arbitrary spin 
state with respect to the conical axis, 
$|{\rm in}>=\alpha|\Uparrow>+\beta|\Downarrow>$,
the right lead transmitted state is
$|rl>=(\alpha t_{\Uparrow\Uparrow}+\beta t_{\Uparrow\Downarrow})|\Uparrow>
+(\alpha t_{\Downarrow\Uparrow}+\beta t_{\Downarrow\Downarrow})|\Downarrow>$
and the left reflected state
$|ll>=(\alpha r_{\Uparrow\Uparrow}+\beta r_{\Uparrow\Downarrow})|\Uparrow>
+(\alpha r_{\Downarrow\Uparrow}+\beta r_{\Downarrow\Downarrow})|\Downarrow>$.
The transmission, reflection coefficients to an outgoing state 
$|{\rm out}>=\gamma|\Uparrow>+\delta|\Downarrow>$ are, 
\begin{equation}
t_{{\rm out,in}}=<{\rm out}|rl>,~~~ r_{{\rm out,in}}=<{\rm out}|ll>.
\label{inout}
\end{equation}

As an example, for the ${\hat y}$ - conical magnet with incoming electron 
spin-$\sigma$  in the $+{\hat z}$ direction, 
$|{\rm in}>=(1/{\sqrt 2})(|\Uparrow>+|\Downarrow>)$, it follows that 
in the potential and diffraction windows,
$|t_{\sigma',\sigma}|^2=|r_{\sigma',\sigma}|^2=1/4$.
We should mention again that these coefficients hold in the large 
$L$ limit (discussed below). 

We can also define the expectation values
of the transmitted-reflected electron spin components $x,y,z$ 
of a left incoming electron in the spin state $|{\rm in}>$ as,
\begin{equation*}
\sigma^{x,y,z}_t=<rl|\sigma^{x,y,z}|rl>,~~~
\sigma^{x,y,z}_r=<ll|\sigma^{x,y,z}|ll>.
\end{equation*}
\noindent

For instance,
in a ${\hat y}$ - conical magnet, an incoming electron with spin 
in the $\pm {\hat y}$ -  direction, the transmission-reflection 
spin compoments are,    

\begin{center}
\begin{tabular}{|c|c|c|}
  \hline 
 $+{\hat y}$ &$\sigma^y_r$ &  $\sigma^y_t$  \\
  \hline
$\epsilon < \epsilon_0$  & 0 & +1  \\ 
  \hline
$\epsilon_{\ne}$ & 0 & +1   \\ 
  \hline
$\epsilon_-< \epsilon < \epsilon_+$  & 0 & +1  \\ 
  \hline
\end{tabular}~~~~
\begin{tabular}{|c|c|c|}
  \hline 
$-{\hat y}$ &$\sigma^y_r$ &  $\sigma^y_t$  \\
  \hline
$\epsilon < \epsilon_0$  & -1 & 0  \\ 
  \hline
$\epsilon_{\ne}$ & 0 & -1   \\ 
  \hline
$\epsilon_-< \epsilon < \epsilon_+$  & +1 & 0  \\ 
  \hline
\end{tabular}
\end{center}
\noindent
In the potential energy region $\epsilon < \epsilon_0$ a
$+{\hat y}$ electron spin sees a negative potential, eq.(\ref{vx}), 
and it is fully transmitted, 
while a $-{\hat y}$ sees a positive and it is fully reflected.
For $\epsilon_{\ne}$ (meaning energy outside the potential  or diffractive 
window) the electron is fully transmitted.
In the diffractive window, an electron with spin parallel to the 
chirality is transmitted while one with spin opposite to the chirality 
is reflected with spin-flip.  

For an electron with spin in the ${\hat x} - {\hat z}$ plane,
\begin{center}
\begin{tabular}{|c|c|c|c|}
  \hline ${\hat x} - {\hat z}$ 
 &$\sigma^y_r$ & $(\sigma^x_t)^2+(\sigma^y_t)^2$ & $\sigma^y_t$  \\
  \hline
$\epsilon < \epsilon_0$ & -1/2 & 0 & +1/2  \\ 
  \hline
$\epsilon_{\ne}$ & 0 & +1 & 0  \\ 
  \hline
$\epsilon_-< \epsilon < \epsilon_+$ & +1/2 & 0 & +1/2  \\ 
  \hline
\end{tabular}
\end{center}
\noindent
This selectivity is understood by considering as the propagation of 
an electron with incoming spin in the ${\hat x} - {\hat z}$ direction 
as a superposition of $\pm {\hat y}$ spin states.
For $\epsilon_{\ne}$ the electron is transmitted with spin in the 
${\hat x} - {\hat z}$ plane at an angle dependent on the 
energy of the incoming electron.

We can discuss analytically the 
diffraction window $Q$ selectivity conditions and size $L$ dependence 
for the spiral ${\hat z}$ - conical 
magnet in the ${\hat x} - {\hat y}$ plane, 
($\theta=0,~~{\vec h}=h(\cos Qy,\sin Qy,0)$), with 
Hamiltonian, 
\begin{equation*}
H=
-\frac{\partial^2}{\partial y^2}-h
\begin{pmatrix}
0& e^{-iQy}\\
e^{+iQy} & 0
\end{pmatrix}.
\end{equation*}
\noindent
The eigenfunctions are of the form,
\begin{equation*}
\Psi=
e^{+iqy}
\begin{bmatrix}
\alpha e^{-iQy/2}\\
\beta e^{+iQy/2}
\end{bmatrix}
\end{equation*}
\noindent
with eigenvalues $\epsilon_{\pm}=q^2+(Q/2)^2\pm
\sqrt{(qQ)^2+h^2}$, which gives for the $\epsilon_-$ branch,
$\alpha/\beta=t+\sqrt{1+t^2},~~t=qQ/h$.
There are also eigenfunctions in the middle of the 
spectrum, Fig.\ref{fig3} with complex wavevectors ${\bar q}$ 
of the form,
\begin{equation*}
\Psi=
e^{-{\bar q}y}
\begin{bmatrix}
{\bar\alpha} e^{-iQy/2}\\
{\bar \beta} e^{+iQy/2}
\end{bmatrix}
\end{equation*}
\noindent
with eigenvalues ${\bar \epsilon}_{\pm}=-{\bar q}^2+(Q/2)^2\pm
\sqrt{h^2-({\bar q}Q)^2}$, which implies,
${\bar \alpha}/{\bar \beta}={\bar t}+\sqrt{1+{\bar t}^2},
~~{\bar t}=i{\bar q}Q/h$.

For the wavevector ${\bar q}=h/Q$ which corresponds to the top of 
the lower complex dispersion, we find $q\simeq \pm Q,~~~p\simeq Q/2$. 
As for the parameters in the Figures $qQ/h >>1 (q>0)$, 
$\alpha/\beta >> 1$ and ${\bar \alpha}/{\bar \beta} \rightarrow i$
so that the wavefunction in the conical magnet is of the form,
\begin{equation*} 
\Psi \simeq A
\begin{bmatrix}
1\\
0
\end{bmatrix}
e^{+iQy/2}+B
\begin{bmatrix}
0\\
1
\end{bmatrix}
e^{-iQy/2}+C
e^{-hy/Q}
\begin{bmatrix}
ie^{-iQy/2}\\
e^{+iQy/2}
\end{bmatrix}.
\end{equation*}
The large $L$ limit corresponds to $e^{-hL/Q} \rightarrow 0$.
By continuity of the wavefunction and the 1st derivative at $y=0,L$ 
we find $|t_{\Uparrow\Uparrow}|=1, r_{\Uparrow\Downarrow}=1$.
Inverting the chirality, $Q\rightarrow -Q$, the up and down spin states 
in the wavefunction are inverted.

\section{Conclusions}
We have presented the generic behavior of 
the conductance of an electron gas in double exchange interaction 
with a classical one dimensional 
conical magnet. It shows a rich behavior, depending on the orientation 
of the fictitious conical magnetic field and its chirality. 
By implementing a modulated magnetic field, we disentangled two 
scattering mechanisms that we assigned as {\it potential} and {\it diffractive}.
They give rise to stepwise integer conductance windows 
in energy and spin selectivity.

It would be interesting to experimentally study the stepwise conductance 
and electron spin selectivity by varying the chemical potential 
e.g. scanning the energy window.
Inserting units in the energy dispersion relation $\epsilon=q^2$ and 
taking $q=Q/2=\pi/\lambda$, we find 
$\epsilon=\hbar^2 (Q/2)^2/2m \sim  36 {\rm eV\cdot {\AA}^2}/\lambda^2$,  
where $\lambda$ is the wavelength of the 
conical magnet measured in $\rm {\AA}$. 
As an example, for $\lambda\simeq 10 {\rm\AA}$ 
we obtain $\epsilon \simeq 0.36$eV 
and for a magnetic exchange coupling constant $J$ 
a fraction of $\epsilon$, 
we obtain reasonable energy parameters for the  
observation of the two separate, potential and diffractive conductance 
windows. In this example, $\epsilon \sim 4000$K, thus considering 
the low temperature limit should not be an issue and 
the $L\rightarrow \infty$ limit is only relevant for obtaining 
sharp conductance steps (in our calculations, the case of $m=64$ is 
shown, $L\sim 64$ nm). 
Furthermore, as the conductance steps are due to the  
robust $O(J)$ gap in the energy spectrum, 
we do not expect disorder/decoherence effects to 
qualitatively alter the overall picture of the conductance.

A rapidly increasing number of synthesized and studied 
chiral compounds \cite{ciss,goehler,cheong}, 
as bulk itinerant helimagnets, dichalcogenides,
chiral metal oxide thin films or chiral chemical, biological molecules,  
could provide platforms for the observation of this conductance pattern. 
To realize the varying field profile, recent progress in 
tailored artificial chiral magnets \cite{levy,tailored} might also be 
promising. 

Last but not least, any chiral system coupling a fictitious field $h$
to the electron spin as in a double exchange interaction (\ref{vx}) 
will show the CISS phenomenon.

\section{Acknowledments}
I would like to acknowledge the financial support and hospitality of the 
Max Planck Institute for Complex Systems in Dresden, where part of this 
work was carried out as well as constructive discussions 
with A. Kl\"umper and C. Hess.

\end{document}